\documentclass[onecolumn,floatfix,prd,aps,10pt]{revtex4-2}
\usepackage[applemac]{inputenc}
\usepackage[T1]{fontenc}
\pdfoutput=1
\usepackage{enumerate}
\usepackage{braket}
\usepackage{amsmath}
\usepackage{subfigure}
\usepackage{lipsum}
\usepackage{amsfonts}
\usepackage{amssymb, mathrsfs}
\usepackage{bm}
\usepackage[pdftex]{graphicx}
\usepackage{footnote}
\usepackage{hyperref}  
\hypersetup{colorlinks=true,allcolors=blue}
\usepackage{hypcap}   
\usepackage{bookmark}
\usepackage{dsfont,mathrsfs,color,url,verbatim,booktabs}

\def\beq{\begin{equation}}
	\def\eeq{\end{equation}}
\def\bsp{\begin{split}}
	\def\esp{\end{split}}
\def\bea{\begin{eqnarray}}
	\def\eea{\end{eqnarray}}
\def\ba{\begin{array}}
	\def\ea{\end{array}}

\def\l.{\left.}
\def\r.{\right.}

\def\part{\partial}
\def\tfrac#1#2{{\textstyle{#1\over #2}}}

\begin{document}
	
	\title{Teleparallel $F(T)$ electromagnetic static spherically symmetric spacetime solutions}
	\author{A. Landry}
	\email{a.landry@dal.ca}
	\affiliation{Department of Mathematics and Statistics, Dalhousie University, Halifax, Nova Scotia, Canada, B3H 3J5}

\begin{abstract}
	We investigate static, spherically symmetric (SS) spacetimes in covariant teleparallel \(F(T)\) gravity in the presence of electromagnetic sources. Starting from the coframe/spin-connection (CSC) pair formalism, we derive the field equations and associated conservation laws, which constrain admissible electromagnetic configurations and reconstructed teleparallel sectors. A general reconstruction procedure is established, allowing the systematic construction of nonlinear teleparallel \(F(T)\) models for arbitrary coframe ans\"atze. Focusing on power-law (PL) configurations, we obtain several classes of exact solutions, including constant-radius, black-hole-like (BH-like), and wormhole-like (WH-like) branches, and analyze their horizon structures, torsion singularities, and stability properties. The inclusion of electromagnetic sources leads to new charged solutions that generalize Reissner--Nordstr\"om (RN) spacetimes and reveal modified near-horizon and asymptotic behaviors. The results are further organized within an invariant classification framework, highlighting the role of torsion in shaping the solution space. Overall, this work provides a unified and covariant approach to the construction and interpretation of physically relevant compact-object, effective cosmological, and regularized strong-field sectors in nonlinear teleparallel gravity, with potential implications for strong-field tests beyond General Relativity (GR).
\end{abstract}

\maketitle

\section{Introduction}

The teleparallel formulation of gravity provides an alternative geometrical description of gravitation in which torsion, rather than curvature, encodes the gravitational interaction. In~this framework, gravity is described by the tetrad (coframe) field and a flat spin--connection (SC), leading to the Teleparallel Equivalent of General Relativity (TEGR), which is dynamically equivalent to Einstein's theory~\cite{Hayashi1979, Aldrovandi2013, Maluf2013, Arcos2004}. The~torsion scalar $T$ plays the role of the Lagrangian density, replacing the Ricci scalar $R$ of standard General Relativity~(GR).

A natural extension of TEGR is obtained by promoting the torsion scalar to an arbitrary function $F(T)$, giving rise to modified teleparallel theories of gravity~\cite{Ferraro2007, Bengochea2009, Linder2010}. These theories have attracted considerable attention as viable alternatives to GR, particularly in the context of cosmology, where they can account for the late-time acceleration of the Universe without invoking dark energy~\cite{Cai2016, Bahamonde2023}. In~addition, $F(T)$ models have been extensively studied in relation to cosmography, large-scale structure formation, and~observational constraints~\mbox{\cite{Capozziello2011,Dent2011,Wu2010,Izumi2013, Li2011, Nesseris2013}}.

Despite their phenomenological success, early formulations of teleparallel $F(T)$ gravity suffered from a lack of local Lorentz invariance, which raised fundamental concerns about their physical consistency~\cite{Li2011}. This issue was later resolved through the development of a fully covariant formulation based on the inclusion of a non-trivial SC~\cite{Krssak2015, Krssak2019, Golovnev2021}. In~this covariant approach, the~pair $(h^{a}{}_{\mu}, \omega^{a}{}_{b\mu})$ ensures both local Lorentz invariance and a consistent geometric interpretation of torsion. This formulation has become the standard framework for modern investigations in teleparallel gravity. {In this context, physically meaningful solutions must be constructed within the covariant coframe/spin-connection (CSC) pair framework to avoid spurious constraints arising from non-invariant tetrad choices. In~non-covariant formulations of teleparallel gravity, the~choice of frame may artificially constrain the admissible solution space through frame-dependent antisymmetric field equations. Such formulations may generate spurious restrictions on spherically symmetric (SS) configurations and obscure the physical interpretation of torsion degrees of~freedom.

The covariant formulation resolves this issue by consistently separating inertial and gravitational contributions through the introduction of a flat SC. In~Einstein--Maxwell systems, this distinction becomes particularly important since improper tetrad choices may generate inconsistent electromagnetic sectors or spurious torsion constraints. Consequently, the~CSC formalism provides the natural covariant framework for constructing physically admissible static SS solutions in nonlinear $F(T)$ gravity.}

Beyond cosmology, an~important line of research concerns the construction and classification of exact solutions in $F(T)$ gravity. In~contrast to GR, in which 
Birkhoff's theorem strongly constrains SS solutions, modified teleparallel theories exhibit an enlarged class of solution space~\cite{DeBenedictis2022, Wang2011, Daouda2012}. In~particular, static and SS configurations have been extensively investigated, including relativistic stars, anisotropic fluids, and~black hole (BH) solutions~\cite{Boehmer2011, Junior2015}. 

A key development in this direction is the invariant classification program of teleparallel geometries, which was initiated in analogy with the Cartan--Karlhede algorithm in GR. Recent works by Coley, and~collaborators have demonstrated that teleparallel spacetimes can be classified using torsion invariants and Cartan scalars, thus providing a powerful tool to distinguish inequivalent geometries~\cite{Coley2020, McNutt2023, Coley2024}. This approach has been further extended to $F(T)$ gravity, where the role of symmetry and invariant structures becomes even more subtle due to the nonlinearity of the field equations. In~this context, recent studies have explored SS and cosmological solutions using invariant methods and covariant formulations~\cite{Landry2024_spherical,Landry2024_fluid,Landry2025_scalar,roberthudsonSSpaper,Bahamonde2019}.

An additional layer of complexity arises when matter fields are included, particularly electromagnetic fields. The~coupling between teleparallel gravity and Maxwell theory leads to an enlarged class of charged solutions that generalize the Reissner--Nordstr\"om spacetime of GR. Several works have investigated charged BHs and electromagnetic configurations in $F(T)$ gravity~\cite{Golovnev2020, Awad2017, Nashed2019}, revealing novel features such as modified horizon structures, deviations from standard asymptotics, and~potential violations of no-go theorems known in GR. However, a~systematic and covariant treatment of Einstein--Maxwell systems in $F(T)$ gravity, especially within the CSC formalism, {remains incomplete.}

From a theoretical perspective, the~interplay between torsion and electromagnetism raises fundamental questions regarding the role of gauge symmetries, conservation laws, and~invariant structures. While Maxwell's equations retain their standard form in curved spacetime, their coupling to torsion can induce nontrivial modifications in the gravitational sector, affecting both the field equations and their solutions. Furthermore, the~presence of electromagnetic sources provides a natural arena to test the consistency and predictive power of modified teleparallel theories beyond vacuum~configurations.

{{Despite significant progress on exact solutions in modified teleparallel gravity, several important aspects remain incompletely understood. In~particular, a~unified covariant treatment simultaneously incorporating electromagnetic sources, conservation laws, invariant classification methods, and~reconstruction procedures has not yet been systematically~developed.} 
	
	While charged solutions and SS geometries have been investigated in several particular contexts, the~interplay between torsion invariants, Maxwell fields, and~admissible nonlinear $F(T)$ sectors remains largely unexplored. Moreover, the~role played by antisymmetric field equations in constraining electromagnetic configurations and reconstructed teleparallel models has not been fully clarified within the invariant classification program.} Motivated by these considerations, the~goal of this work is to develop a systematic and covariant analysis of SS solutions in $F(T)$ gravity with electromagnetic sources. We adopt the CSC formalism to ensure full local Lorentz invariance and construct the corresponding field equations in the presence of a Maxwell field, as done for other types of sources in Refs.~\mbox{\cite{Landry2024_spherical,Landry2024_fluid,Landry2025_scalar,roberthudsonSSpaper}}. Particular attention is devoted to the role of symmetry, invariant classification, and~the structure of solution~space. 

In addition, we aim to bridge the gap between formal developments and physically relevant models by deriving explicit classes of solutions and analyzing their properties. This includes the investigation of power-law (PL) ans\"atze, charged configurations, and~the behavior of conservation laws in the absence or presence of electric charge. {In particular, the~reconstruction procedure allows the reduced field equations to be transformed into ordinary differential equations for the unknown function \(F(T)\).} We also discuss the implications of our results for observational signatures and possible deviations from~GR.

{
To summarize concretely, the~main contributions of the present work~are:
\begin{itemize}
	\item We derive the covariant Einstein--Maxwell field equations for static SS teleparallel geometries using the CSC~formalism.
	
	\item We obtain explicit conservation law solutions for radial electric, magnetic, and~mixed electromagnetic sectors compatible with spherical~symmetry.
	
	\item We construct a closed-form reconstruction procedure that allows for the determination of admissible nonlinear $F(T)$ models for PL coframe ans\"atze.
	
	\item We extend the teleparallel invariant classification program to electromagnetic sectors in covariant $F(T)$ gravity.
	
	\item We analyze the corresponding horizon structures, torsion singularities, and~stability conditions associated with reconstructed solution~branches.
	
	\item We investigate wormhole-like (WH-like) sectors and discuss how nonlinear torsion contributions may effectively support WH-like geometries by shifting the NEC-violating contribution to the effective torsion sector rather than imposing it directly on the physical Maxwell source.
\end{itemize}

{From} 
 a physical perspective, these solutions provide a unified framework to investigate modified compact objects, generalized Reissner--Nordstr\"om (RN) geometries, effective cosmological sectors, and~WH-like configurations within nonlinear teleparallel $F(T)$ gravity. In~addition, the~reconstructed models may lead to observable deviations from GR through modifications of horizon structures, strong-field lensing, BH shadows, and~quasi-normal mode~spectra.

The paper is organized as follows. In~Section~\ref{sect2}, we review the covariant formulation of teleparallel $F(T)$ gravity, introduce the CSC formalism, and~derive the Einstein--Maxwell field equations together with the associated conservation laws. In~Sections~\ref{sect3}--\ref{sect5}, we specialize to static SS configurations and investigate constant-radius, BH-like, and~WH-like sectors using reconstruction and invariant classification techniques. We then analyze the corresponding stability conditions, horizon structures, and~singularity properties. Finally, in~Section~\ref{sect6}, we summarize our results, discuss the physical limitations of the present approach, and~outline possible future~developments.

}


\section{Teleparallel Field Equations, Maxwell Sector and Invariant~Classification}\label{sect2}
\unskip

\subsection{Teleparallel $F(T)$ Gravity~Framework}\label{sect21}

We consider the covariant formulation of teleparallel $F(T)$ gravity, where the fundamental variables are the coframe $h^a_{\ \mu}$ and a flat SC $\omega^a_{\ b\mu}$ ensuring local Lorentz invariance~\cite{Krssak2015,Cai2016,Bahamonde2023,Bahamonde2023b,Landry2024_spherical}.

The action is given  {by} 
\begin{equation}
	S = \int d^4x \left[ \frac{h}{2\kappa} F(T) - \frac{1}{4} F_{\mu\nu}F^{\mu\nu} + A_\mu J^\mu \right],
\end{equation}
where $h = \det(h^a_{\ \mu})$. {We adopt the metric signature \((- ,+,+,+)\) and use natural units \(c=G=1\) throughout.}

The torsion tensor, superpotential and torsion scalar are defined as
\begin{align}
	T^a_{\ \mu\nu} &= \partial_\mu h^a_{\ \nu} - \partial_\nu h^a_{\ \mu}
	+ \omega^a_{\ b\mu} h^b_{\ \nu} - \omega^a_{\ b\nu} h^b_{\ \mu}, \\
	S_a^{\ \mu\nu} &= \frac{1}{2} \left(T^{\mu\nu}_{\ \ a} + T^{\nu\mu}_{\ \ a} - T_a^{\ \mu\nu}\right)
	- h_a^{\ \nu} T^{\lambda\mu}_{\ \ \lambda} + h_a^{\ \mu} T^{\lambda\nu}_{\ \ \lambda}, \\
	T &= \frac{1}{2} T^a_{\ \mu\nu} S_a^{\ \mu\nu}. \label{torsionscalar4}
\end{align}

Variation of the action yields the field equations
\begin{equation}\label{FEsgeneral}
	\kappa \Theta^\mu_{\ a} =
	h^{-1} F_T \partial_\nu (h S_a^{\ \mu\nu})
	+ F_{TT} S_a^{\ \mu\nu} \partial_\nu T
	+ \frac{F}{2} h^\mu_{\ a}
	- F_T (T^b_{\ a\nu} + \omega^b_{\ a\nu}) S_b^{\ \mu\nu}.
\end{equation}
 {In} 
 terms of symmetric and antisymmetric tangent-space components, we decompose {{Equation}
~\eqref{FEsgeneral}} \cite{Landry2024_spherical}:
\begin{align}
	\kappa\,\Theta_{\left(ab\right)} =& F_T \overset{\ \circ}{G}_{ab}+F_{TT}\,S_{\left(ab\right)}^{\;\;\;\mu}\,\partial_{\mu} T+\frac{g_{ab}}{2}\,\left[F-T\,F_T\right],  \label{1001a}
	\\
	0 =& F_{TT}\left(T\right)\,S_{\left[ab\right]}^{\;\;\;\mu}\,\partial_{\mu} T. \label{1001b}
\end{align}
{%
	Equations~\eqref{1001a} and \eqref{1001b} follow from the decomposition of {Equation}~\eqref{FEsgeneral} into symmetric and antisymmetric components with respect to the tangent-space indices. Equation \eqref{1001b} represents the antisymmetric sector arising from local Lorentz covariance~constraints.

	The structure of these field equations reveals that nonlinear torsion corrections couple directly to both the symmetric gravitational sector and the antisymmetric coframe constraints. As~a consequence, the~admissible electromagnetic configurations become strongly restricted in comparison with Einstein--Maxwell~theory.

	 The decomposition into symmetric and antisymmetric parts plays a central role in teleparallel $F(T)$ gravity. The~symmetric field equations determine the dynamical gravitational part, while the antisymmetric field equations constrain the admissible CSC pairs and strongly restrict nonlinear torsion configurations. In~particular, the~antisymmetric parts impose severe conditions on the existence of nontrivial electromagnetic configurations in modified teleparallel gravity. The~restrictive role of the antisymmetric sector in nonlinear teleparallel $F(T)$ gravity has been emphasized in several studies, particularly regarding the admissibility of SS coframes and charged configurations~\cite{Krssak2015,Golovnev2020,Bahamonde2019,Coley2020,Landry2024_spherical}.}

\subsection{Coframe/Spin-Connection and Torsion~Structure}\label{sect22}

In order to construct exact solutions, it is convenient to adopt a coframe-based approach. In~teleparallel gravity, the~fundamental variables are the tetrad (coframe) $h^a_{\ \mu}$ and the flat SC $\omega^a_{\ b\mu}$, which together encode both inertial and gravitational effects~\cite{Krssak2015,Cai2016,Landry2024_spherical}.

We consider a general static, SS spacetime described by the CSC pair~\cite{Landry2024_spherical,Landry2024_fluid,Landry2025_scalar,roberthudsonSSpaper}:
\begin{eqnarray}
	h^a_{\ \mu} &=& \mathrm{diag}\left(A_1(r),\, A_2(r),\, A_3(r),\, A_3(r)\sin\theta \right), \label{coframe}
	\\
	\omega_{233} &=& \omega_{244} = \frac{\delta}{A_3(r)},~ \omega_{344} = - \frac{\cot(\theta)}{A_3(r)} \label{spincon}
\end{eqnarray}
which leads to the metric
\begin{equation}
	ds^2 = -A_1^2(r)\, dt^2 + A_2^2(r)\, dr^2 + A_3^2(r)\, d\Omega^2.
\end{equation}
{The CSC pair \eqref{coframe} and \eqref{spincon} corresponds to an orthonormal covariant coframe/spin-connection configuration. The~diagonal coframe is adopted for simplicity and compatibility with static spherical symmetry. The~choice of gauge must be supplemented by an appropriate CSC pair to ensure vanishing curvature and preserve local Lorentz invariance~\cite{Krssak2015,Coley2020,Landry2024_spherical}. In~the covariant formulation, this guarantees that inertial effects are properly separated from gravitational contributions. In~short, the~SC encodes inertial contributions associated with the choice of gauge and frame, thereby ensuring the local Lorentz covariance of the~theory.

}

The torsion tensor is defined by
\begin{equation}\label{torsionten}
	T^a_{\ \mu\nu} = \partial_\mu h^a_{\ \nu} - \partial_\nu h^a_{\ \mu}
	+ \omega^a_{\ b\mu} h^b_{\ \nu} - \omega^a_{\ b\nu} h^b_{\ \mu}.
\end{equation}
For the above CSC pair, the~torsion scalar takes the general form
\begin{equation}\label{1001c}
	T = T\left(A_1, A_2, A_3, A_1', A_3'\right),
\end{equation}
which depends explicitly on radial derivatives. {Substituting the CSC pair \eqref{coframe} and \eqref{spincon} into Equation~\eqref{torsionten} and contracting with the superpotential gives the torsion scalar structure summarized in Equation~\eqref{1001c}.} This scalar encodes all gravitational degrees of freedom in $F(T)$ gravity~\cite{Cai2016,Bahamonde2023,Bahamonde2023b}.

\subsection{Solutions of Maxwell Equations and Conservation~Laws}\label{sect23}

In this section, we derive the general solutions of the covariant Maxwell equations together with the associated conservation laws in a static SS spacetime. These results provide the electromagnetic sector that will be coupled to the $F(T)$ gravitational field~equations.

We consider the covariant Maxwell equations~\cite{Jackson1999,LandauLifshitz1975,HehlObukhov2003}
\begin{equation}
	\overset{\circ}{\nabla}_\nu F^{\mu\nu} = J^\mu,
\end{equation}
together with the Bianchi identity
\begin{equation}\label{bianchi}
	\overset{\circ}{\nabla}_{[\lambda} F_{\mu\nu]} = 0.
\end{equation}
The electromagnetic energy-momentum tensor is given by
\begin{equation}
	\Theta_{\mu\nu} = F_{\mu\alpha} F_{\nu}^{\ \alpha}
	- \frac{1}{4} g_{\mu\nu} F_{\alpha\beta} F^{\alpha\beta}, 
\end{equation}
and {the corresponding conservation law reads,}
\begin{equation}\label{eqn16}
	\overset{\circ}{\nabla}_\nu \Theta^{\mu\nu} = 0.
\end{equation}
{These equations are consistent with the current conservation law and ensure compatibility with the gravitational field equations.
}

Taking the divergence of the Maxwell equations and using the antisymmetry of $F^{\mu\nu}$, one obtains the conservation of the four-current~\cite{Jackson1999,LandauLifshitz1975,HehlObukhov2003}
\begin{equation}\label{current}
	\overset{\circ}{\nabla}_\mu J^\mu = 0.
\end{equation}
{Under static spherical symmetry, the~Maxwell sector reduces to purely radial conservation laws. This strongly constrains transverse electromagnetic modes and naturally favors Coulomb-like electric or magnetic configurations compatible with the antisymmetric teleparallel constraints. This reflects the compatibility between spherical symmetry and the gauge structure of Maxwell theory in non-flat~spacetime.

 The covariant formulation of Maxwell electrodynamics in curved and torsionful backgrounds follows the standard gauge-invariant construction developed in classical field theory and relativistic electrodynamics~\cite{Jackson1999,LandauLifshitz1975,HehlObukhov2003}.}

\begin{enumerate}

\item \textbf{{Current conservation law solution:}
} {Substituting the static SS metric determinant into Equation~\eqref{current}, the~current conservation law reduces to}
\begin{equation}
	\frac{1}{h} \frac{d}{dr} \left( h J^r \right) = 0,
\end{equation}
which admits the general solution
\begin{equation}
	J^r(r) = \frac{J_0}{A_1(r) A_2(r) A_3^2(r)},
\end{equation}
where $J_0$ is a constant. In~most physical situations involving static configurations, one sets $J^r = 0$, implying the absence of radial current~flow.

Similarly, the~temporal component satisfies
\begin{equation}
	J^t = \rho(r),
\end{equation}
with $\rho(r)$ constrained by global charge~conservation.

\item \textbf{{Radial Electric Field:}} For a purely radial electric field, the~only non-vanishing component is
\begin{equation}\label{eqn21}
	F_{tr} = -F_{rt} = E(r).
\end{equation}

{Using the purely radial electric ansatz \eqref{eqn21}, the~Maxwell equations simplify to}
\begin{equation}
	\frac{1}{h} \frac{d}{dr} \left(h F^{tr} \right) = J^t.
\end{equation}

In vacuum ($J^\mu = 0$), this integrates to
\begin{equation}
	F^{tr} = \frac{Q}{A_1 A_2 A_3^2},
\end{equation}
leading to the solution
\begin{equation}
	E(r) = \frac{Q}{A_3^2(r)}.
\end{equation}

In the presence of a charge density $\rho(r)$, the~solution generalizes to
\begin{equation}
	F^{tr}(r) = \frac{1}{A_1 A_2 A_3^2} \int^r \rho(\tilde r)\, A_1 A_2 A_3^2 \, d\tilde r.
\end{equation}

\item \textbf{{Radial Magnetic Field:}} A radial magnetic field corresponds to the angular component
\begin{equation}\label{eqn26}
	F_{\theta\phi} = B(r)\, A_3^2(r)\sin\theta.
\end{equation}

{Applying the Bianchi identity \eqref{bianchi} to the magnetic ansatz \eqref{eqn26}, we obtain}
\begin{equation}
	\frac{d}{dr} \left( B(r) A_3^2(r) \right) = 0,
\end{equation}
which integrates to
\begin{equation}
	B(r) = \frac{B_0}{A_3^2(r)},
\end{equation}
where $B_0$ is a constant magnetic~charge.

This solution automatically satisfies current conservation since it is purely topological in~origin.

\item \textbf{{Transverse Electromagnetic Fields:}} Transverse fields involve components such as $F_{t\theta}$, $F_{t\phi}$, $F_{r\theta}$, $F_{r\phi}$. {For transverse configurations, the~Maxwell equations reduce to the generic conservation form }
\begin{equation}
	\frac{1}{h} \frac{d}{dr} \left(h F^{\mu r} \right) = J^\mu.
\end{equation}

Using current conservation, the~general solution takes the form
\begin{equation}
	F^{\mu r}(r) = \frac{C^\mu}{A_1 A_2 A_3^2}
	+ \frac{1}{A_1 A_2 A_3^2} \int^r J^\mu(\tilde r)\, A_1 A_2 A_3^2 \, d\tilde r,
\end{equation}
where $C^\mu$ are~constants.

{These configurations represent formal transverse electromagnetic modes; however, in~the admissible covariant SS branches considered below, the~antisymmetric field equations and conservation laws strongly constrain or eliminate them unless additional symmetry-breaking sectors are introduced.}

\item \textbf{{Mixed Electromagnetic Configurations:}} The most general configuration combines electric, magnetic, and~transverse components:
\begin{align}
	F_{tr} &= E(r), \\
	F_{\theta\phi} &= B(r) A_3^2 \sin\theta, \\
	F^{\mu r} &= \frac{C^\mu}{A_1 A_2 A_3^2}.
\end{align}

The corresponding solutions are
\begin{align}
	E(r) &= \frac{Q}{A_3^2(r)} + \text{source terms}, \\
	B(r) &= \frac{B_0}{A_3^2(r)}, \\
	F^{\mu r} &= \frac{C^\mu}{A_1 A_2 A_3^2} + \text{source terms}.
\end{align}
These solutions satisfy both Maxwell equations and current conservation identically. {The electric and magnetic sectors remain formally dual under $Q \leftrightarrow B_0$ at the level of Maxwell equations.}

\item \textbf{{Summary of conservation-law solutions in vacuum:}} The combined Maxwell and conservation laws admit the following general structure in pure vacuum ($J^\mu =0$)
\begin{itemize}
	\item Electric field: $E(r) = \frac{Q}{A_3^2(r)}$.
	
	\item Magnetic field: $B(r) = \frac{B_0}{A_3^2(r)}$.
	
	\item Transverse fields: $F^{\mu r} = \frac{C^\mu}{A_1 A_2 A_3^2}$.
\end{itemize}
These results are independent of the specific form of $F(T)$ and provide the fundamental electromagnetic sector used in the construction of exact solutions in the following sections in the vacuum~sector.

\end{enumerate}


\subsection{Static Electromagnetic Field~Equations}\label{sect24}

The SFEs and torsion scalar expressions are derived from Equation \eqref{1001a} \cite{Landry2024_spherical,Landry2024_fluid,Landry2025_scalar,roberthudsonSSpaper}.

\subsubsection{$A_3=c_0=$ Constant}

For \(A_3=c_0=\mathrm{constant}\), Equations~\eqref{1001a} and \eqref{1001c} give:
\begin{align}
	\kappa \rho_{em}=&-\tfrac{1}{2}\left[F-TF_T\right] -\frac{2\delta\partial_r\left(F_T\right)}{A_2c_0} +\frac{F_T}{c_0^2} , \label{2203a}
	\\
	\kappa\,P_r =& \tfrac{1}{2}\left[F-TF_T\right]-\frac{F_T}{c_0^2}  , \label{2203b}
	\\
	-\kappa\,P_r=& \tfrac{1}{2}\left[F-TF_T\right]+\frac{\partial_r\left(F_T\right)}{A_2}\left[\frac{\delta}{c_0}+\left(\frac{A_1'}{A_1A_2}\right)\right]
	+F_T\left[\frac{A_1''}{A_2^2A_1}-\left(\frac{A_1'}{A_1A_2}\right)\left(\frac{A_2'}{A_2^2}\right)\right]  , \label{2203c}
	\\
	T(r)=& -2\left(\frac{\delta}{c_0}\right)\left(\frac{\delta}{c_0}+\frac{2\,A_1'}{A_1\,A_2}\right). \label{2203d}
\end{align}
{Equation~\eqref{2203d} follows directly from substituting the constant-radius condition \(A_3=c_0\) into the general torsion scalar expression \eqref{1001c}. This branch is analogous to generalized Nariai or Bertotti--Robinson sectors. Note that if $A_1=A_{10}=$ constant, we obtain from Equation~\eqref{2203d} that $T=-\frac{2}{c_0^2}=$ constant: TdS-like or GR-like branches~\cite{TdSpaper}. The~constant torsion contribution effectively behaves as a cosmological vacuum sector.} The conservation law solutions in vacuum~are:
\begin{itemize}
	\item Electric field: $E(r) = E_0=$ constant.
	
	\item Magnetic field: $B(r) = {B_0}=$ constant.
	
	\item Transverse fields: $F^{\mu r} = \frac{\tilde{C}^\mu}{A_1 A_2}$.
\end{itemize}

{{The} 
 electromagnetic energy density is $\rho=-P_r=P_t = \frac{E_0^2 + B_0^2}{2} = \mathrm{const}$ and conservation laws are trivially satisfied. The~electromagnetic density remains constant in this branch and therefore contributes as an effective constant source term in the reduced field equations, although~the pressure structure remains~anisotropic.

The antisymmetric field equations impose strong restrictions on the admissible CSC pairs. In~particular, mixed electromagnetic configurations require constant torsion, transverse modes are excluded, and~only radial electric and magnetic fields remain physically admissible within the covariant SS sector~\cite{Krssak2015,Golovnev2020}. }

\subsubsection{$A_3=r$}

For $A_3=r$, Equations~\eqref{1001a} and \eqref{1001c} give:\vspace{6pt}
{\small\begin{align}
	\kappa \rho_{em}=&-\tfrac{1}{2}\left[F-TF_T\right] -\frac{2\partial_r\left(F_T\right)}{A_2}\left(\frac{\delta}{r}+\frac{1}{A_2r}\right) +F_T\left[2\left(\frac{A_1'}{A_1A_2}\right)\left(\frac{1}{A_2r}\right)-\left(\frac{1}{A_2r}\right)^2+\frac{1}{r^2}\right] , \label{2204a}
	\\
	\kappa\,P_r =& \tfrac{1}{2}\left[F-TF_T\right]+F_T\left[2\left(\frac{A_1'}{A_1A_2}\right)\left(\frac{1}{A_2r}\right)+\left(\frac{1}{A_2r}\right)^2-\frac{1}{r^2}\right]  , \label{2204b}
	\\
	-\kappa\,P_r=& \tfrac{1}{2}\left[F-TF_T\right]+\frac{\partial_r\left(F_T\right)}{A_2}\left[\frac{\delta}{r}+\left(\frac{A_1'}{A_1A_2}\right)+\left(\frac{1}{A_2r}\right)\right]
	\nonumber\\
	&\quad+F_T\left[\frac{A_1''}{A_2^2A_1}-\left(\frac{A_1'}{A_1A_2}\right)\left(\frac{A_2'}{A_2^2}\right)+\left(\frac{A_1'}{A_1A_2}\right)\left(\frac{1}{A_2r}\right)-\left(\frac{1}{A_2r}\right)\left(\frac{A_2'}{A_2^2}\right)\right]  , \label{2204c}
	\\
	T(r)=& -2\left(\frac{\delta}{r}+\frac{1}{A_2\,r}\right)\left(\frac{\delta}{r}+\frac{1}{A_2\,r}+\frac{2\,A_1'}{A_1\,A_2}\right). \label{2204d}
\end{align}}

The conservation law solutions in vacuum~are:
\begin{itemize}
	\item Electric field: $E(r) = \frac{Q}{r^2}$.
	
	\item Magnetic field: $B(r) = \frac{B_0}{r^2}$.
	
	\item Transverse fields: $F^{\mu r} = \frac{C^\mu}{A_1 A_2 r^2}$.
\end{itemize}

The energy density is $\rho(r)=-P_r=P_t = \frac{Q^2 + B_0^2}{2 r^4}$. {The Maxwell sector therefore reproduces the standard RN Coulomb scaling. Substituting the Coulomb scaling into the conservation law \eqref{eqn16}, we obtain}
\begin{equation}\label{eqn45}
	\frac{d\rho}{dr} + \frac{4}{r}\rho = 0, \quad\Rightarrow\quad	\rho(r) \sim \frac{1}{r^4}.
\end{equation}
{Equation \eqref{eqn45} reproduces the expected inverse quartic decay of electromagnetic energy density in SS symmetry. The~antisymmetric field equations imply strong restrictions on admissible CSC pairs: nonlinear \(F(T)\) sectors require non-constant torsion \(T=T(r)\), while transverse electromagnetic modes vanish in the admissible vacuum branch (\(\tilde C^\mu=0\)) \cite{Krssak2015,Golovnev2020}.

The same Maxwell scaling and antisymmetric field equation restrictions will also be used in the WH-like sector of Section~\ref{sect5}.}

\subsection{teleparallel Invariant~Classification}\label{sect25}

To systematically classify the solutions of $F(T)$ gravity, we adopt the invariant formalism developed by Coley and collaborators, which extends Cartan's equivalence method to teleparallel geometries~\cite{Landry2024_spherical,Landry2024_fluid,Landry2025_scalar,roberthudsonSSpaper,Coley2009,Olver1995}. This approach relies on scalar invariants constructed from the torsion tensor and its covariant derivatives, providing a coordinate- and frame-independent characterization of spacetime~geometries.

The fundamental torsion invariants are given by
\begin{equation}\label{eqn46}
	I_1 = T, \quad
	I_2 = T_{\mu\nu\rho}T^{\mu\nu\rho}, \quad
	I_3 = S_{\mu\nu\rho}S^{\mu\nu\rho}, \quad
	I_4 = \nabla_\mu T \nabla^\mu T,
\end{equation}
which encode the essential geometric and dynamical properties of the torsion~field.

Based on these invariants, the~admissible $F(T)$ models can be classified into distinct~families:

\begin{itemize}
	\item \textbf{{TEGR class:}
} $F(T)=\alpha T+\beta$, corresponding to constant torsion $T=T_0$ and equivalent to GR with an effective cosmological constant.
	\item \textbf{{PL class:}} {$F(T)=T^n$, leading to algebraic torsion invariants and scale-invariant~solutions.}
	\item \textbf{{LOG class:}} $F(T)=T^n \ln(T/T_0)$, generating logarithmic corrections.
	\item \textbf{{EXP class:}} $F(T)=e^{\lambda T}$, typically associated with strong-field regimes.
	\item \textbf{{COMP class:}} combinations of the above, yielding richer phenomenology.
\end{itemize}

A complete local geometric characterization requires the Cartan invariant set
\begin{equation}
	\mathcal{I} = \{T, \nabla T, \nabla^2 T, \dots\},
\end{equation}
which uniquely determines the spacetime up to local Lorentz transformations. {In practice, these invariant sets provide a coordinate-independent method to distinguish inequivalent teleparallel branches generated by the reconstruction procedure.} Two solutions are equivalent if and only if their invariant sets coincide {\cite{Coley2009,Coley2020}. Invariant classification methods based on Cartan equivalence and scalar torsion invariants have proven particularly useful for identifying geometrically distinct teleparallel sectors independently of coordinate or gauge choices~\cite{Coley2020,McNutt2023,Olver1995}. The~Cartan invariant set provides a locally complete geometric characterization of teleparallel spacetimes.}

Physically, constant-torsion solutions correspond to effective cosmological constant sectors, PL models describe scale-invariant regimes, while EXP and LOG forms capture strong-field and quantum-like corrections. Deviations from GR arise primarily from the nonlinear dependence on $T$, while the Maxwell sector retains its standard gauge structure~\cite{Golovnev2020,Awad2017,Nashed2019}.

{
Having established the covariant Einstein--Maxwell framework and the associated conservation laws, we now focus on static SS field equations and investigate exact reconstructed teleparallel $F(T)$ solutions. 

}


\section{Exact Solutions for \boldmath{$A_3 = c_0$} in the~Vacuum}\label{sect3}

{In this section, we investigate the first class of exact reconstructed teleparallel $F(T)$ functions generated by static SS coframe configurations. The~objective is to determine how nonlinear torsion corrections modify the horizon structure, singularity behavior, invariant classification, and~dynamical stability of charged Einstein--Maxwell solutions in covariant \(F(T)\) gravity.}

\subsection{$F(T)$ Vacuum~Solutions}

{

	In this first step, we consider the vacuum electromagnetic case defined by \(A_3(r)=c_0=\text{const}\) and \(J^\mu=0\), corresponding to static configurations. This regime is particularly relevant for near-horizon geometries and effective vacuum states in modified teleparallel~gravity.
	
	The simplest class of solutions corresponds to TEGR-like models of the form
\begin{equation}
		F(T)=\alpha T + \beta,
	\end{equation}
	which always admit consistent solutions. In~this case, the~constant \(\beta\) effectively absorbs the electromagnetic vacuum energy, leading to a theory equivalent to GR with an effective cosmological~constant.
	
	Constant-torsion configurations, defined by \(T=T_0\), similarly reduce to GR-like solutions. These correspond to vacuum branches where torsion behaves as an effective cosmological~sector.
	
	Beyond these trivial configurations, nonlinear teleparallel models give rise to richer structures. PL models of the form \(F(T)=T^n\) generate scale-invariant solutions and introduce nontrivial modifications to the gravitational sector. LOG extensions,
\begin{equation}
	F(T)=T + \alpha T \ln\left(\frac{T}{T_0}\right),
	\end{equation}
	provide mild corrections and are typically associated with quantum-inspired modifications. EXP models,
\begin{equation}
	F(T)=T + \alpha e^{\lambda T},
	\end{equation}
	lead to strong-field corrections and may admit constant-torsion vacuum configurations, although~they are generally more sensitive to~perturbations.
	
	More generally, COMP models combining these contributions allow for multi-branch structures and richer phenomenology, interpolating between different invariant classes of teleparallel geometries. A~typical example is:
\begin{equation}
	F(T)=T + \alpha T^n + \beta e^{\lambda T} .
	\end{equation}

Unlike GR, nonlinear torsion corrections allow several inequivalent vacuum branches for the same symmetry sector. The~corresponding analytical branches and their geometric interpretations are summarized in Table~\ref{table1}. Table~\ref{table1} also indicates that stability is closely tied to the behavior of $F_{TT}$.}

\begin{table}[ht]
	\begin{tabular}{ccccc}
		\toprule
		\textbf{Class} & \boldmath{$F(T)$} & \textbf{Torsion} & \textbf{Stability} & \textbf{Interpretation} \\
		\midrule
		TEGR & $\alpha T+\beta$ & variable & Stable & GR + $\Lambda$ \\
		\midrule
		Const-T & constant & constant & Stable & vacuum branch \\
		\midrule
		PL & $T^n$ & variable & Conditionally stable & scaling vacuum \\
		\midrule
		LOG & $T+\alpha T\ln T$ & variable & Stable & quantum-like vacuum \\
		\midrule
		EXP & $T+\alpha e^{\lambda T}$ & constrained & Sensitive & strong-field vacuum \\
		\midrule
		COMP & mixed & variable & Model dependent & rich vacuum \\
		\bottomrule
	\end{tabular}
		\caption{{The} teleparallel invariant classification table of $A_3=c_0=$ constant class of~solutions.}
	\label{table1}
\end{table}
\unskip

\subsection{Closed-Form Power-Law Reconstruction, Classification and~Stability}

We consider the constant $A_3(r)=c_0$ sector and adopt the PL ansatz $A_1(r)=a_0 r^a$ and $A_2(r)=b_0 r^b$ with constants $a_0,b_0>0$. {We also consider a non-vacuum solution by adding a $\rho_{\rm em}\neq 0$ electromagnetic contribution.} From Equation~\eqref{2203d}, the~torsion scalar reads
\begin{equation}
	T(r)=T_0 + T_1 r^{-(b+1)},
\end{equation}
where $T_0=-\frac{2\delta^2}{c_0^2}$ and $T_1=-\frac{4a\delta}{c_0 b_0}$.

For $b\neq -1$,
\begin{equation}\label{inversion}
	r^{-(b+1)}=\frac{T-T_0}{T_1}.
\end{equation}

{
	
	The Equation~\eqref{inversion} is valid provided \(T(r)\) remains monotonic in the domain of interest, ensuring a well-defined mapping between the radial coordinate and the torsion scalar. This condition restricts the admissible parameter space \((a,b)\) and guarantees the consistency of the reconstruction~procedure.
	
	To obtain explicit reconstructed teleparallel $F(T)$ solutions, we specialize to PL coframe configurations characterized by two independent scaling parameters \((a,b)\). These parameters control both the asymptotic behavior of the lapse function and the scaling structure of the torsion scalar $T(r)$.

	 Substituting the PL coframe ansatz into the reduced field equations and expressing the radial coordinate as a function of the torsion scalar allows the reconstruction equations to be reduced to an ordinary differential equation for the unknown function $F(T)$. This procedure provides a systematic mechanism to identify admissible nonlinear teleparallel models compatible with the assumed symmetry structure. Using the inversion relation \eqref{inversion}, all radial powers appearing in the reduced field equations can be rewritten as powers of \(T-T_0\). Substituting this result into Equations~\eqref{2203a}--\eqref{2203d} gives}
\begin{equation}\label{eqn50}
	(T-T_0)F_{TT} + \gamma(a,b)\,F_T = \Lambda(a,b,T),
\end{equation}
{where $\gamma(a,b)=\frac{b+1-2a}{b+1}$ and $\Lambda(a,b,T)=\frac{c_0^2}{2\delta^2}(b+1)\,\kappa \rho_{\rm em}(T)$. Equation~\eqref{eqn50} has the structure of a generalized Euler-type differential equation in the shifted variable $(T - T_0)$, whose coefficients depend only on the scaling parameters $(a, b)$ and the electromagnetic source sector. The~reconstruction equation should be interpreted as a consistency condition selecting admissible effective teleparallel sectors within the assumed symmetry ansatz. This structure explains why PL, LOG, and~COMP solutions naturally emerge within the reconstruction~scheme.

 The reconstruction equation shows that nonlinear torsion corrections are directly controlled by the interplay between the scaling parameters \((a,b)\) and the electromagnetic source term encoded in \(\Lambda(a,b,T)\). This highlights the role of symmetry and conservation laws in selecting admissible teleparallel~models.

The homogeneous part of Equation~\eqref{eqn50} leads to
\begin{equation}\label{eqn55}
F(T)=C_1 (T-T_0)^{n} + C_2.
\end{equation} 
For a generic source term \(\Lambda(a,b,T)\), we set $F_T=Y(T)$, $\Lambda(a,b,T)=\tilde{\Lambda}(a,b)\rho_{\rm em}(T)$ and then Equation~ \eqref{eqn50} becomes:
\begin{equation}\label{eqn52}
	Y_{T} + \frac{\gamma(a,b)}{(T-T_0)}\,Y = \tilde{\Lambda}(a,b)\frac{\rho_{\rm em}(T)}{(T-T_0)}.
\end{equation}
Solving Equation~\eqref{eqn52} with the integrating factor \((T-T_0)^{\gamma(a,b)}\), we obtain
{\small\begin{equation}\label{gensol}
	F(T)=\tilde{\Lambda}(a,b)\int_T\,dT'(T'-T_0)^{n-1}\left[\int_{T'}\,dT''\,\rho_{\rm em}(T'')(T''-T_0)^{-n}\right] + C_1 (T-T_0)^{n} + C_2,
\end{equation}}%
where $n=\frac{2a}{b+1}$. Equation \eqref{gensol} explicitly demonstrates that the reconstructed function \(F(T)\) inherits the scaling properties of both the torsion scalar and the electromagnetic sector, leading to a hierarchy of admissible nonlinear models. This gives three representative~cases:
\begin{itemize}
\item $\rho_{\rm em}=\rho_{\rm em\,0}\,(T-T_0)^p$ ($p \neq \left\lbrace-1,\,0 \right\rbrace$):
\begin{equation}\label{casep}
	F(T)=\frac{\Lambda(a,b)}{(p+\gamma)(p+1)}\,(T-T_0)^{p+1} + C_1 (T-T_0)^{n} + C_2.
\end{equation}	
	
\item $p=0$ ($\rho_{\rm em}=$ constant):
\begin{equation}\label{casep0}
	F(T)=\frac{\Lambda(a,b)}{\gamma}(T-T_0) + C_1 (T-T_0)^{n} + C_2 .
\end{equation}

\item $p=-1$:
\begin{equation}\label{casepmin1}
	F(T)=\frac{\Lambda(a,b)}{(\gamma-1)}\,\ln(T-T_0) + C_1 (T-T_0)^{n} + C_2.
\end{equation}
\end{itemize}

{Physically, the~parameter \(a\) in Equations \eqref{gensol}--\eqref{casepmin1} primarily governs the horizon structure through the scaling behavior of the lapse function, while the parameter \(b\) controls the near-core behavior of torsion invariants and therefore the singularity structure of the reconstructed~geometry.

The classification of reconstructed solutions follows directly from Equations~\eqref{gensol}--\eqref{casepmin1}. TEGR-like configurations arise when \(a = 0\) or \(b = -1\), corresponding to constant-torsion sectors or effective cosmological branches. PL models are obtained for generic values of
\begin{equation}
n=\frac{2a}{b+1}
\end{equation}
with \(p\neq \{-1,0\}\), leading to scale-invariant modifications of the gravitational~sector.

LOG regimes emerge when \(2a = b + 1\) or \(p = -1\), where nonlinear torsion corrections introduce LOG scaling behavior. EXP sectors correspond formally to the limit \(b + 1 \to 0\) or \(p \to \infty\), and~are associated with strongly nonlinear torsion dynamics. Finally, COMP models arise from superpositions of these contributions and lead to multi-scale teleparallel~geometries.

More physically, the~scaling parameters \((a,b)\) therefore encode the competition between horizon formation and torsion regularization. While the parameter \(a\) primarily controls the redshift structure and the existence of compact trapped regions, the~parameter \(b\) governs the ultraviolet behavior of torsion invariants near the geometric core. From~a broader perspective, these reconstructed solutions may be interpreted as effective geometric phases of teleparallel gravity. Each class corresponds to a distinct balance between torsion dynamics, electromagnetic contributions, and~symmetry constraints, leading to qualitatively different compact-object configurations. This mechanism explains the emergence of scale-invariant and multi-branch teleparallel $F(T)$ models.

}

{

\subsection{Critical Points and Singular~Structures}

The reconstructed teleparallel sectors may develop critical behavior either through geometric divergences near \(r=0\) or through degeneracies associated with the nonlinear torsion sector. In~particular, singular configurations may arise whenever torsion invariants diverge or when the effective coupling factor \(F_T\) approaches~zero.

Thus, the~singularity analysis is performed at the level of torsion invariants rather than only through the metric functions. For~\(b>-1\), the~torsion scalar generally diverges near the origin, leading to BH-like geometries with central singularities. The~critical branch \(b=-1\) corresponds to softer LOG behavior, while configurations with \(b<-1\) may regularize the torsion sector and generate nonsingular compact~cores.

Configurations satisfying \(F_T=0\) formally require special attention, as~they correspond to a vanishing effective gravitational coupling, potentially signaling strong-coupling regimes or a breakdown of the effective teleparallel~description.

Additional degeneracies may occur when \(F_{TT}=0\), signaling the transition between dynamically distinct teleparallel~branches.

These singular behaviors can be systematically characterized using the torsion invariants introduced in Section~\ref{sect25}, which provide a coordinate-independent diagnostic of the geometric structure of the~solutions.

}

\subsection{{Stability Analysis and Physical~Interpretation}}

{
	
	The stability properties of the reconstructed teleparallel models can be analyzed by considering linear perturbations around a background torsion configuration \(T = T_0\). In~this framework, the~effective scalar degree of freedom is characterized by the ratio
\begin{equation}
	m_{\text{eff}}^2 \sim \frac{F_T}{F_{TT}},
	\end{equation}
	which provides a leading-order diagnostic for ghost and tachyonic instabilities up to model-dependent normalization factors. This expression should be interpreted as a leading-order scalar-sector diagnostic rather than a complete perturbative stability~criterion.
	
	Stable configurations correspond to \(F_T > 0\) and \(F_{TT} > 0\), ensuring a positive effective mass and the absence of pathological modes. Conversely, negative values of \(F_{TT}\) signal the presence of tachyonic instabilities and dynamical instability of the torsion~sector.
	
	Although this criterion captures the leading-order behavior, a~complete stability analysis would require the study of coupled perturbations involving both torsion and metric degrees of~freedom.
	
	In the constant-radius regime, the~reconstructed solutions correspond to limiting geometries such as Nariai-type spacetimes, near-horizon configurations of charged BHs, and~effective cosmological vacua. Although~not BHs themselves, these solutions describe limiting geometries of generalized Reissner--Nordstr\"om spacetimes~\cite{Awad2017,Nashed2019}.

For the PL reconstructed branch, substituting Equation~\eqref{eqn55} into the definitions of \(F_T\) and \(F_{TT}\) gives}
\begin{equation}
	F_T = C_1 n (T-T_0)^{n-1} + \frac{\Lambda}{\gamma},\qquad 	F_{TT} = C_1 n(n-1)(T-T_0)^{n-2}.
\end{equation}

The absence of ghost and tachyonic instabilities also requires $F_T > 0$ and $F_{TT} > 0$. {The ghost-free condition is therefore
\begin{equation}\label{eqn64}
C_1 n (T-T_0)^{n-1}+\frac{\Lambda}{\gamma}>0,
\end{equation}
while the absence of tachyonic modes requires
\begin{equation}\label{eqn65}
C_1 n(n-1)(T-T_0)^{n-2}>0.
\end{equation}

	These inequalities can be recast in terms of the effective exponent \(n = \frac{2a}{b+1}\), which provides a more transparent classification of stability regimes. Configurations with \(n>1\) correspond to potentially stable modified-gravity regimes, provided \mbox{Equations~\eqref{eqn64} and \eqref{eqn65}} hold on the physical branch considered. The~critical case \(n = 1\) reproduces the TEGR limit. Values in the range \(0 < n < 1\) lead to infrared instabilities, and~negative values of \(n\) indicate pathological strong-coupling behavior. More concretely:}	
\begin{itemize}	
	\item \textbf{{Marginal (LOG):}
}
\begin{equation}
		2a = b+1 \quad \Rightarrow \quad F_{TT}\to 0.
	\end{equation}
	
	\item \textbf{{Unstable regime:}}
\begin{equation}
		0 < \frac{2a}{b+1} < 1.
	\end{equation}
	
	\item \textbf{{Pathological regime:}}
\begin{equation}
		\frac{2a}{b+1} < 0
		\quad (\text{ghost or tachyon instabilities}).
	\end{equation}
\end{itemize}

Thus, the~effective exponent \(n=2a/(b+1)\) controls both the invariant class and the leading scalar-torsion stability~properties.

In summary, the~reconstructed teleparallel $F(T)$ solutions exhibit a rich interplay between torsion dynamics, electromagnetic contributions, and~geometric structure. Depending on the choice of parameters, the~solutions may interpolate between GR-like vacuum configurations, modified compact objects, and~potentially regularized geometries supported by nonlinear torsion effects. These results emphasize that teleparallel \(F(T)\) gravity provides a unified framework in which geometry, matter coupling, and~stability properties are intrinsically linked through the torsion~sector.

}


\section{Exact Solutions for \boldmath{$A_3=r$} in the~Vacuum}\label{sect4}

{We now consider the physically most relevant sector \(A_3(r)=r\), which corresponds to the standard areal-radius gauge for static SS geometries. In~contrast with the constant-radius sector of Section~\ref{sect3}, this class contains BH-like configurations, charged compact objects, and~their nonlinear teleparallel generalizations~\cite{Cai2016,Bahamonde2023}. In~vacuum, the~Maxwell sector reduces to the usual Coulomb scaling, \(E(r)\sim Q/r^2\) and \(B(r)\sim P/r^2\), while the nonlinear dependence of \(F(T)\) modifies the gravitational field equations through torsion corrections.}

\subsection{BH--Like~Solutions}

\begin{figure}[ht]
	\hspace{-0.6cm}\includegraphics[width=0.85\textwidth]{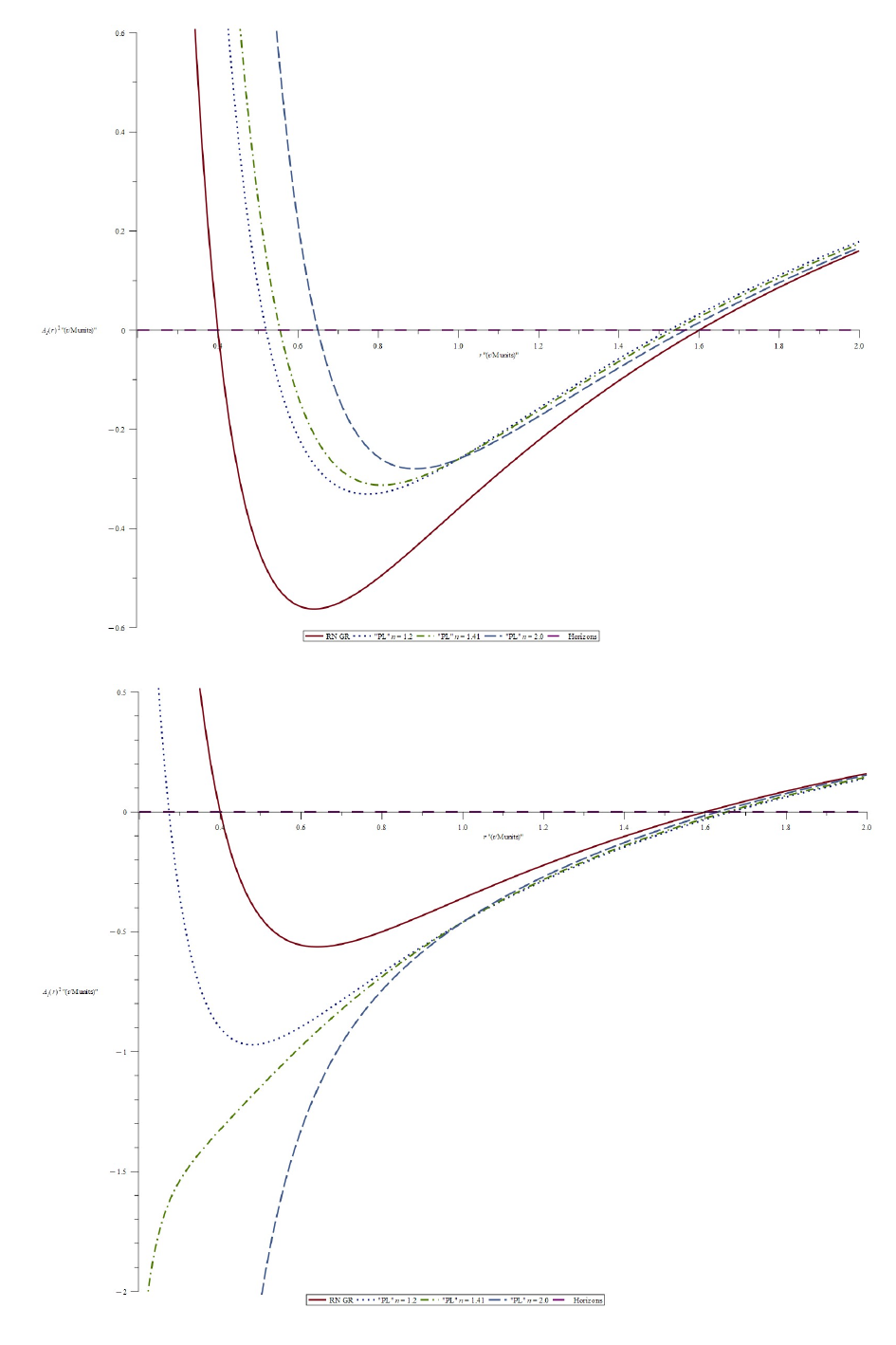}\\\vspace{-4pt}	  
	\caption{{Dimensionless} 
 plots of \(A_1^2(r/M)\) versus \(r/M\) for PL cases $n=1.2$, $1.41$ and $2.0$, $\alpha=0.1$ (\mbox{\textbf{top plot}}) and $\alpha=-0.1$ (\textbf{bottom plot}) with $M=1$ and $Q=0.8$. All quantities are normalized with respect to the mass scale \(M\), with~\(c=G=1\).}
	\label{figure1}
\end{figure}

\begin{figure}[ht]
	\hspace{-0.6cm}\includegraphics[width=0.95\textwidth]{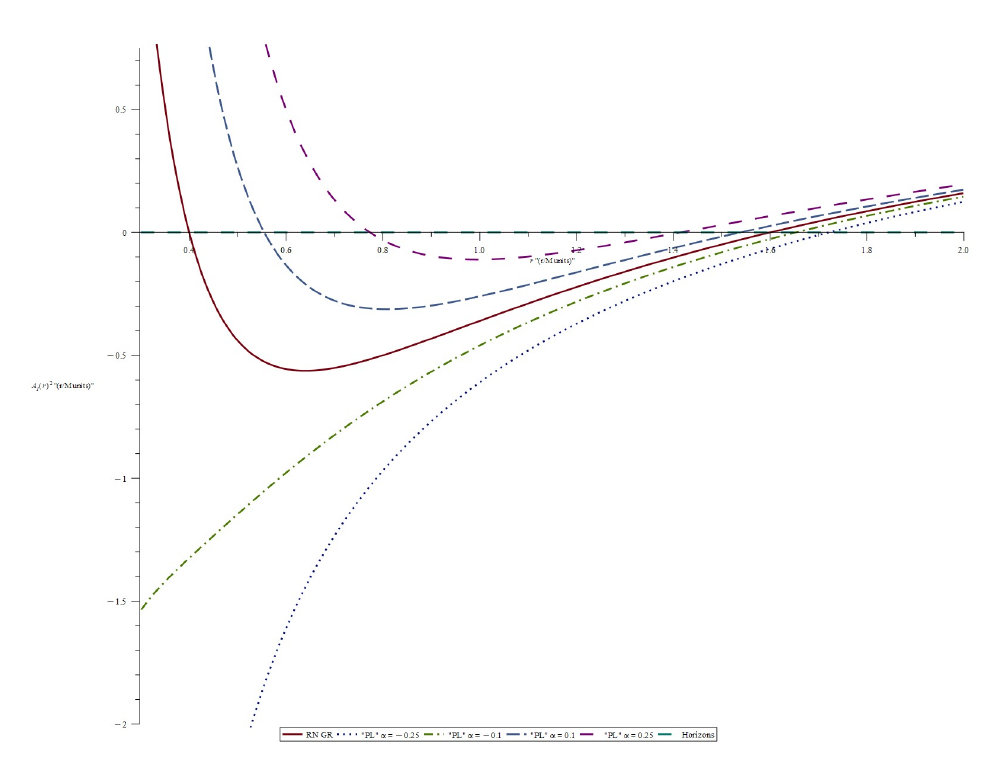}\\\vspace{-22pt}   
	\caption{{Dimensionless} 
 plots of \(A_1^2(r/M)\) versus \(r/M\) for PL cases $n=1.41$, $\alpha=-0.25$ to $0.25$ with $M=1$ and $Q=0.8$. All quantities are normalized with respect to the mass scale \(M\), with~\(c=G=1\).}
	\label{figure2}
\end{figure}

\begin{figure}[ht]
		\hspace{-0.6cm}\includegraphics[width=0.95\textwidth]{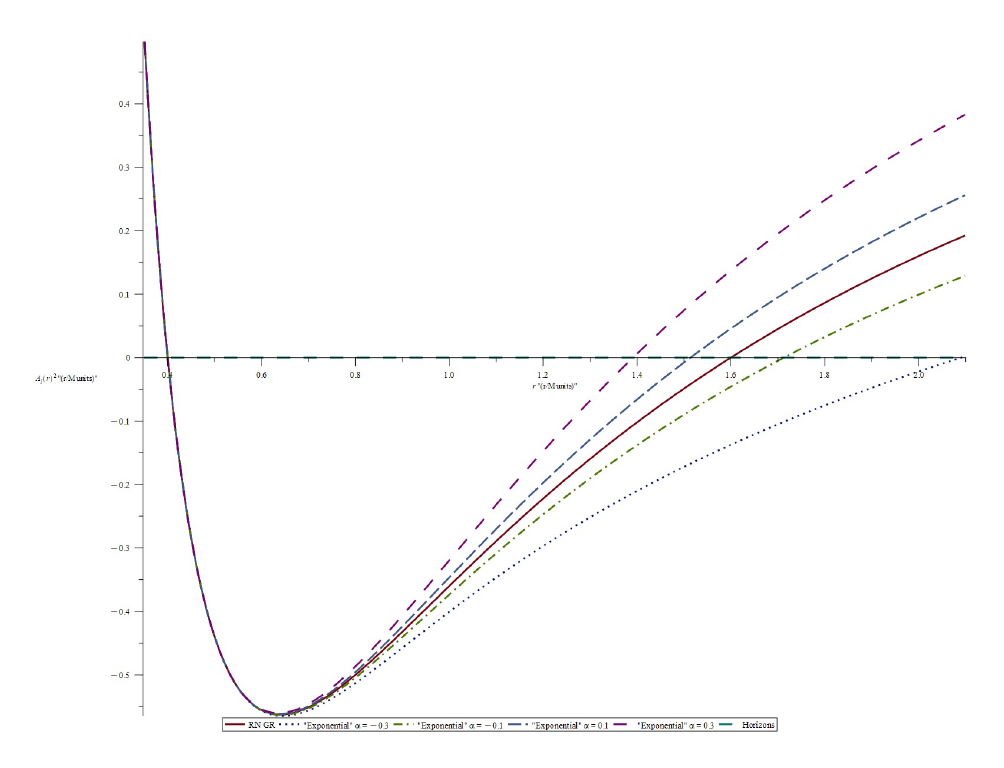}\\\vspace{-26pt}   
	\caption{{Dimensionless} 
 plots of \(A_1^2(r/M)\) versus \(r/M\) for EXP case $\lambda=1$, $T=-\frac{2}{r^2}$, $\alpha=-0.3$ to $0.3$ with $M=1$ and $Q=0.8$. All quantities are normalized with respect to the mass scale \(M\), with~\(c=G=1\).}
	\label{figure3}
\end{figure}

{

In the TEGR limit \(F(T)=T\), the~standard RN-type geometry is recovered,
\begin{equation}
A_1^2=A_2^{-2}=1-\frac{2M}{r}+\frac{Q^2+P^2}{r^2},
\end{equation}
with horizon radii
\begin{equation}
r_{\pm}=M\pm\sqrt{M^2-(Q^2+P^2)}.
\end{equation}
This solution provides the reference GR branch against which nonlinear teleparallel corrections can be~compared.

For PL models of the form
\begin{equation}
F(T)=T+\alpha T^n,
\end{equation}
the lapse function acquires short-distance corrections of the schematic form
\begin{equation}
A_1^2(r)\sim 1-\frac{2M}{r}+\frac{Q^2}{r^2}+\alpha r^{-2n}.
\end{equation}
The exponent \(n\) controls the strength of torsion correction near the compact core, while the sign and magnitude of \(\alpha\) determine whether the correction enhances or weakens the effective gravitational potential, as in Refs.~\cite{Awad2017,Nashed2019,Junior2015}. Figure~\ref{figure1} illustrates the effect of the PL exponent \(n\) on the lapse function \(A_1^2(r)\) for fixed \(\alpha=\pm 0.1\), while Figure~\ref{figure2} shows how varying \(\alpha\) shifts the horizon structure for fixed \(n=1.41\). The~solid curve corresponds to the GR/RN limit \(\alpha=0\). 

LOG models,
\begin{equation}
F(T)=T+\alpha T\ln\left(\frac{T}{T_0}\right),
\end{equation}
generate mild infrared or asymptotic corrections, whereas EXP models,
\begin{equation}
F(T)=T+\alpha e^{\lambda T},
\end{equation}
can produce stronger nonlinear modifications in the high-torsion regime. These branches may modify the horizon structure, effective energy conditions (ECs), and~strong-field observables relative to the RN geometry. Figure~\ref{figure3} displays the corresponding EXP correction, showing how the parameter \(\alpha\) modifies the near-core behavior while preserving the RN branch as a reference~solution.

The resulting invariant classes are summarized in Table~\ref{table2}, which organizes the \(A_3=r\) sector according to the reconstructed \(F(T)\) model, geometric interpretation, horizon structure, and~stability behavior. Physically, Table~\ref{table2} highlights that nonlinear torsion corrections can either preserve, deform, or~entirely remove horizon structures. In~particular, EXP and COMP models provide natural candidates for torsion-regularized BH-like configurations within the present framework. The~zeros of the plotted lapse functions indicate candidate horizon locations, while changes in the number and position of these zeros illustrate how nonlinear torsion corrections deform the RN horizon structure.
}

At a more structural level, the~nonlinear function \(F(T)\) does not modify the Maxwell sector directly but~instead deforms the effective gravitational potential through torsion-induced corrections in the field equations. As~a consequence, different functional forms of \(F(T)\) map to distinct geometric deformations of the lapse function \(A_1^2(r)\), which control horizon formation, photon sphere structure, and~strong-field~observables.
\unskip


\vspace*{0.2cm}
\begin{table}[ht]
	\begin{tabular}{cccccc}
		\toprule
		\textbf{Class} & \boldmath{$F(T)$} & \textbf{Geometry} & \textbf{Horizon} & \textbf{Stability} & \textbf{Notes} \\
		\midrule
		TEGR & $T$ & RN & 2 & Stable & GR limit \\
		\midrule
		PL & $T+\alpha T^n$ & Deformed RN & 1--2 & Cond. stable & short-distance corr. \\
		\midrule
		LOG & $T+\alpha T\ln T$ & Asymp. mod. & 1--2 & Stable & IR corr. \\
		\midrule
		EXP & $T+\alpha e^{\lambda T}$ & Regular BH & 0--2 & Sensitive & core regularization \\
		\midrule
		COMP & mixed & rich & multi & Model dep. & phase structure \\
		\bottomrule
	\end{tabular}
		\caption{The teleparallel invariant classification table of $A_3=r$ solutions.}
	\label{table2}
\end{table}

In contrast with GR, where Birkhoff's theorem enforces the uniqueness of static SS vacuum solutions, the~nonlinear structure of \(F(T)\) gravity allows multiple inequivalent branches for the same symmetry class. This non-uniqueness is directly reflected in the variety of reconstructed solutions presented in this~section.

\subsection{Extended Power-Law Reconstruction, Invariant Classification, Horizons and Singularities for $A_3=r$}

{
	
While the previous subsection focused on specific functional forms of $F(T)$, we now generalize the analysis by performing a systematic reconstruction based on the PL coframe ansatz, allowing for a unified treatment of multi-scale torsion effects. For~the PL coframe ansatz $A_1(r)=a_0 r^a$ and $A_2(r)=b_0 r^b$, the~torsion scalar obtained from Equations~\eqref{2204a}--\eqref{2204d} takes the multi-scale form
\begin{equation}
T(r)=T_0 r^{-2}+T_1 r^{-(b+2)}+T_2 r^{-2(b+1)},
\end{equation}
where
\begin{equation}
T_0=-2,\qquad T_1=\frac{4(1+a)}{b_0},\qquad
T_2=-\frac{2(1+a)^2}{b_0^2}.
\end{equation}
Unlike the \(A_3=c_0\) sector, the~areal-radius case contains several competing radial scalings. This multi-scale torsion structure naturally leads to COMP reconstruction branches and richer invariant~classes.

Beyond simple PL reconstructions, the~multi-scale structure of \(T(r)\) allows several COMP nonlinear models. A~representative double PL branch is
\begin{equation}
F(T)=\alpha (T-T_*)^{n_1}+\beta (T-T_*)^{n_2}+\gamma,
\end{equation}
with
\begin{equation}
n_1=\frac{2a}{b+2},\qquad n_2=\frac{a}{b+1}.
\end{equation}
Log-corrected branches may be written as
\begin{equation}
F(T)=\alpha (T-T_*)^n\left[1+\eta\ln(T-T_*)\right]+\beta T,
\end{equation}
while EXP-power hybrid models take the form
\begin{equation}
F(T)=\alpha (T-T_*)^n+\beta e^{\lambda T}.
\end{equation}
More general rational and running-index models can also be constructed, for~example
\begin{equation}
F(T)=\frac{\alpha (T-T_*)^n}{1+\xi (T-T_*)^m},
\end{equation}
and
\begin{equation}
F(T)=\alpha (T-T_*)^{n(r(T))},\qquad
n(r(T))=\frac{2a}{b+2}+\Delta n\, r^{-\sigma}(T).
\end{equation}
These branches correspond to generalized Invariant COMP invariant classes and encode multi-scale torsion~corrections.

The lapse function associated with the PL ansatz is
\begin{equation}
A_1^2(r)=a_0^2 r^{2a}.
\end{equation}
It should be emphasized that this expression corresponds to the leading 
behavior of the lapse function within the PL ansatz and must be supplemented by the full reconstructed solution to determine the exact causal structure. Candidate horizon locations are determined by the zeros of the full lapse function. The~leading PL expression \(A_1^2(r)=a_0^2r^{2a}\) does not by itself generate a finite-radius horizon. Horizon formation therefore requires the full reconstructed lapse function, where the leading PL behavior is supplemented by RN-like and torsion-induced corrections:
\begin{equation}
A_1^2(r_h)=a_0^2r_h^{2a}+\Delta_{\rm torsion}(r_h)=0,\qquad r_h>0.
\end{equation}
The interplay between these terms determines the number and nature 
of horizons. More generally, the~horizon structure is controlled by the competition 
between the GR-like term and the nonlinear torsion~corrections.

The vanishing or divergence of \(A_1^2(r)\) determines the qualitative causal behavior of the geometry. The~exponent \(a\) controls the redshift structure of the geometry. In~particular, \(a<0\) leads to a decreasing lapse function toward the origin, which is a necessary (but not sufficient) condition for BH horizon formation. The~existence of actual horizons requires matching with subleading terms in the full reconstructed solution. For~\(a\geq0\), no finite-radius zero of the lapse function appears within this simple PL ansatz, and~the geometry is better interpreted as horizonless or naked, depending on the behavior of the torsion invariants. The~case \(a=-1\) reproduces the scaling behavior associated with TEGR/RN-type~sectors.

The singularity structure is controlled primarily by the exponent \(b\), which determines the ultraviolet behavior of the torsion invariants. Near~the origin, the~leading scalings are
\begin{equation}
T\sim r^{-2},\qquad 
I_2=T_{\mu\nu\rho}T^{\mu\nu\rho}\sim r^{-2(b+2)},\qquad
I_3=S_{\mu\nu\rho}S^{\mu\nu\rho}\sim r^{-4(b+1)}.
\end{equation}
{The asymptotic structure further constrains the physical admissibility 
of the solutions. As~$r \to \infty$, the~torsion scalar behaves as 
$T \sim r^{-2}$, ensuring asymptotic flatness for admissible $F(T)$ models. Near~the core ($r \to 0$), the~dominant contribution depends on $b$: $T \sim r^{-2(b+1)}$, leading to divergent, critical, or~regularized 
behavior depending on the sign of $b+1$.}

For \(b>-1\), the~torsion invariants generically diverge as \(r\to0\), indicating a central singularity analogous to the RN core. The~critical case \(b=-1\) corresponds to a softer singular branch, while \(b<-1\) may regularize the torsion sector and generate regular BH-like cores. These critical and divergence points directly address the singularity structure of the lapse and torsion invariants as illustrated in Figure~\ref{figure5}. The~torsion profiles can be correlated with the lapse deformations shown in Figures~\ref{figure1}--\ref{figure3}: divergent torsion for \(b>-1\) corresponds to stronger short-distance corrections, whereas softened torsion behavior for \(b<-1\) supports milder near-core modifications. From~an observational perspective, these torsion-induced deformations may affect strong-field signatures such as photon sphere radii, BH shadows, and~quasi-normal mode spectra, providing potential avenues to constrain the parameter \(b\) through astrophysical~observations. 

\begin{figure}[ht]
\hspace{-0.6cm}\includegraphics[width=0.95\textwidth]{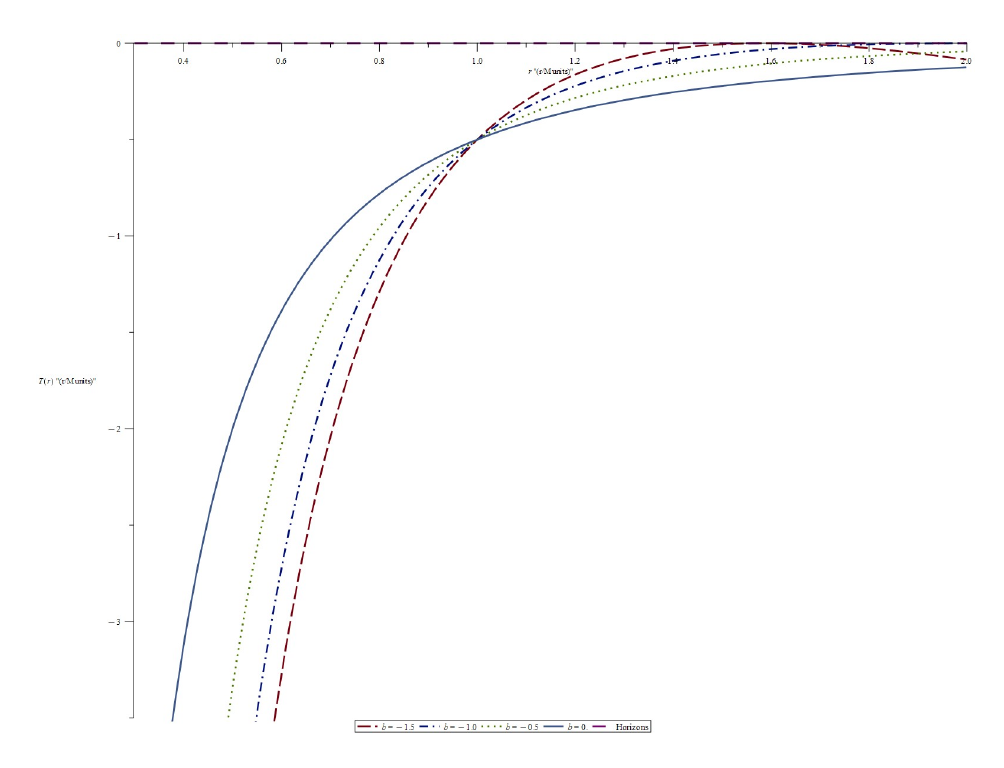}\\\vspace{-26pt}    
	\caption{{Dimensionless} 
 torsion profile \(T(r/M)\) as a function of \(r/M\) for different values of \(b\), illustrating the transition between singular \((b>-1)\), critical \((b=-1)\), and~regularized \((b<-1)\) torsion-core behaviors. Natural units \(c=G=1\) are~used.}
	\label{figure5}
\end{figure}

Within the invariant classification scheme, the~TEGR branch reproduces the usual charged BH singularity at \(r=0\). PL models preserve this singular behavior whenever \(b>-1\), whereas EXP and COMP branches may soften or regularize the torsion invariants depending on the dominant nonlinear contribution. Thus, regularization is not automatic in nonlinear \(F(T)\) gravity; it occurs only in restricted regions of parameter space where the torsion invariants remain~finite.

The effective energy-momentum tensor induced by nonlinear torsion $T(r)$ 
can be defined through the modified field equations, allowing one 
to test ECs. In~particular, regularized branches (\(b< -1\)) may effectively violate classical ECs through the torsion sector, thereby mimicking exotic-matter~contributions.

As in the constant-radius sector, a~leading-order stability diagnostic is obtained from
\[
m_{\rm eff}^2\sim \frac{F_T}{F_{TT}}.
\]
For COMP models,
\[
F_T=\sum_i \alpha_i n_i (T-T_*)^{n_i-1}+\lambda\beta e^{\lambda T},
\]
and
\[
F_{TT}=\sum_i \alpha_i n_i(n_i-1)(T-T_*)^{n_i-2}+\lambda^2\beta e^{\lambda T}.
\]
Stable branches require \(F_T>0\) and \(F_{TT}>0\), which typically selects exponents \(n_i>1\) together with positive EXP contribution \(\beta\lambda^2>0\). The~marginal case \(n_i=1\) corresponds to the TEGR-like limit, while \(0<n_i<1\) leads to infrared instabilities. The~most physically interesting regularizing stable regime is obtained when \(b<-1\) and \(n_i>1\), since this simultaneously softens the torsion core and avoids tachyonic scalar-torsion~modes.

The parameter space $(a, b)$ can therefore be interpreted as defining 
a phase diagram of teleparallel compact objects, separating singular, 
critical, and~regularized geometries. The~space \((a,b)\) separates naturally into four geometric regimes. When \(a<0\) and \(b>-1\), the~solutions describe BH-like geometries with central torsion singularities. When \(a<0\) and \(b<-1\), nonlinear torsion effects may soften the core and produce torsion-regularized BH-like configurations. For~\(a>0\) and \(b>-1\), the~absence of horizons together with divergent torsion invariants leads to naked singular geometries. Finally, \(a>0\) and \(b<-1\) corresponds to regular horizonless sectors. Therefore, the~pair \((a,b)\) controls the interplay between horizon formation, torsion singularities, invariant classification, and~dynamical stability. Taken together, Figures~\ref{figure1}--\ref{figure5} illustrate how the parameters $(a, b)$ and the functional form of $F(T)$ jointly control both the causal 
structure (via $A_1^2$) and the torsion sector (via $T(r)$), providing a unified picture of teleparallel BH~geometries.

For a negative effective cosmological constant, the~reconstructed charged solutions asymptotically approach RN--AdS geometries modified by nonlinear torsion corrections. In~this regime, the~lapse function behaves as
\[
A_1^2(r)\sim 1-\frac{2M}{r}+\frac{Q^2}{r^2}
-\frac{\Lambda_{\rm eff}}{3}r^2+\Delta_T(r),
\]
where \(\Delta_T(r)\) denotes nonlinear torsion corrections. This asymptotic structure shows that nonlinear torsion corrections act as an effective radial-dependent deformation of the cosmological term, potentially leading to deviations from standard AdS asymptotics at intermediate scales. This explicitly displays the AdS branch and clarifies how the effective cosmological constant enters the charged teleparallel BH sector. The~comparison with the GR and RN--AdS lapse functions is illustrated in Figure~\ref{figure4} for representative PL teleparallel~corrections.
\begin{figure}[ht]
\hspace{-0.6cm}\includegraphics[width=0.95\textwidth]{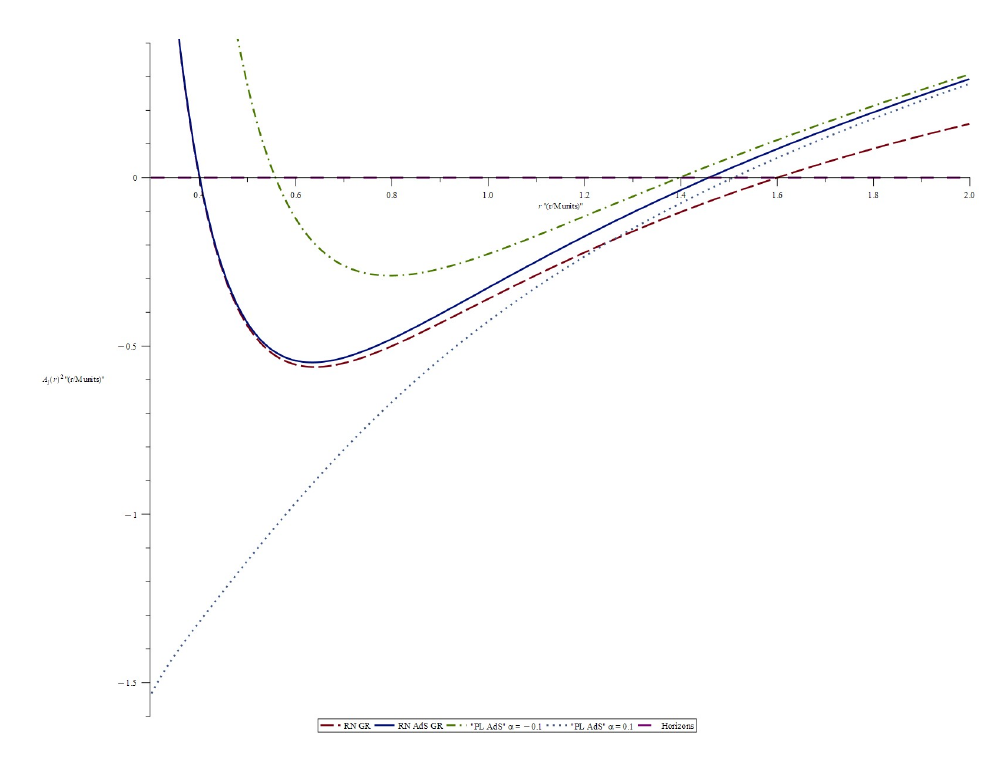}\\\vspace{-26pt}   
	\caption{{Dimensionless} 
 plots of \(A_1^2(r/M)\) versus \(r/M\) for AdS PL cases $n=1.41$, $\Lambda_{\rm eff}=-0.1$ $\alpha=-0.1$ to $0.1$ with $M=1$ and $Q=0.8$. All quantities are normalized with respect to the mass scale \(M\), with~\(c=G=1\).}
	\label{figure4}
\end{figure}
\unskip

\subsection{{Stability Analysis and Physical~Interpretation}}

The stability properties of the \(A_3=r\) sector are governed by the same leading-order scalar-torsion criterion used in Section~\ref{sect3},
\begin{equation}
m_{\rm eff}^2\sim \frac{F_T}{F_{TT}}.
\end{equation}
Stable reconstructed solutions require
\begin{equation}
F_T>0,\qquad F_{TT}>0,
\end{equation}
corresponding respectively to the absence of ghost-like and tachyonic modes. In~this sense, TEGR remains stable because it does not introduce additional nonlinear torsion degrees of freedom. PL models are stable in the regime \(n>1\), LOG corrections are typically marginal or stable near the reference scale \(T_0\), while EXP models are more sensitive to perturbations due to their strong nonlinear dependence on \(T\). These deviations may, in~principle, be probed through observational signatures such as shadow size distortions, deviations in light deflection, or~shifts in quasi-normal mode spectra relative to GR~predictions.

Physically, stable reconstructed contributions correspond to nonlinear torsion sectors capable of supporting compact charged geometries without developing runaway scalar-torsion instabilities. Unstable branches, by~contrast, may generate infrared pathologies or tachyonic growth of effective torsional~modes.

The \(A_3=r\) sector is therefore the most relevant one for compact-object phenomenology. It contains charged teleparallel BHs, deformed RN geometries, possible torsion-regularized BH-like configurations, and~AdS-like charged solutions. The~interplay between torsion and electromagnetic fields may lead to observable deviations in horizon structure, photon spheres, gravitational lensing, BH shadows, and~quasi-normal mode~spectra.

Overall, the~$A_3 = r$ sector reveals a rich landscape of teleparallel 
compact-object solutions, where horizon structure, singularity 
resolution, and~stability are governed by the interplay between 
torsion scaling and electromagnetic contributions. This highlights 
the potential of nonlinear teleparallel $F(T)$ gravity to extend the classical BH paradigm beyond~GR.

}


\section{Wormhole-like Solutions in the~Vacuum}\label{sect5}

{

\subsection{Geometric Conditions for Wormhole~Configurations}

In contrast with BH-like configurations analyzed in Section~\ref{sect4}, WH-like geometries are not characterized by horizon formation but by the existence of a minimal 
areal radius (throat), controlled by the shape function $b(r)$. In~the areal-radius gauge \(A_3=r\), traversable WH geometries can be described 
using the Morris--Thorne parametrization~\cite{MorrisThorne1988}:
\begin{equation}
A_2^{-2}(r)=1-\frac{b(r)}{r}, \qquad A_1^2(r)=e^{2\Phi(r)}.
\end{equation}
The throat radius \(r_0\) is defined by
\begin{equation}
b(r_0)=r_0, \qquad b'(r_0)<1,
\end{equation}
while traversability requires the absence of horizons,
\begin{equation}
\Phi(r_0)<\infty.
\end{equation}

In covariant \(F(T)\) gravity, these geometric conditions must be supplemented by 
constraints arising from the Equations~\eqref{1001a} and \eqref{1001b} \cite{Krssak2015,Krssak2019,Coley2020}. 
In particular, the~antisymmetric sector expressed by Equation \eqref{1001b} restricts admissible CSC pairs and forbids spurious frame choices. The~torsion scalar \(T(r)\), defined in Equation \eqref{torsionscalar4}, provides an invariant characterization 
of the geometry. Regular WH configurations require
\begin{equation}
T(r_0)<\infty, \qquad \partial_r T(r_0)<\infty,
\end{equation}
ensuring that no torsion singularity develops at the~throat.

Thus, WH geometries in \(F(T)\) gravity are determined by a combined set of 
metric, torsional, and~coframe compatibility conditions, extending the standard GR 
analysis~\cite{Bahamonde2023,Bahamonde2023b}. {Equivalently, in~terms of the metric coefficient \(A_2(r)\), the~shape function satisfies
\begin{equation}
	A_2^{-2}(r)=1-\frac{b(r)}{r}.
\end{equation}
Together with \(A_1^2(r_0)=e^{2\Phi(r_0)}\neq0\), these conditions distinguish WH-like throats from BH horizons. The~previous conditions on $b(r_0)$ and $b'(r_0)$ guarantee the existence of a local throat, but~they are not sufficient to prove global traversability or geodesic completeness. Additional asymptotic conditions, such as asymptotic flatness or matching to an external region, must also be~imposed.

Substituting the Morris--Thorne ansatz into the symmetric field equations \eqref{1001a}, evaluated at \(r=r_0\), gives the effective throat constraint
\begin{equation}
\left.(\rho_{\rm eff}+P_{r,\rm eff})\right|_{r_0} =\frac{b'(r_0)-1}{\kappa r_0^2}
+\Delta_T ,
\end{equation}
where \(\Delta_T\) collects the nonlinear torsion contributions. This shows that, unlike 
in GR, the~flaring-out condition is not controlled only by matter sources, but~also by 
the torsion sector. The~antisymmetric field equations impose
\begin{equation}
F_{TT}S^\mu{}_{[ab]}\partial_\mu T=0,
\end{equation}
which requires the contraction between the antisymmetric superpotential and the torsion-gradient sector to vanish. This may occur through the TEGR branch \(F_{TT}=0\), constant-torsion configurations, or~restricted admissible CSC pairs. This demonstrates that, unlike GR, the~throat condition is not purely geometric but depends explicitly on torsion~dynamics.

These geometric conditions can be translated into invariant constraints using the classification of Section~\ref{sect25}, avoiding coordinate-dependent~ambiguities.

}

\subsection{Reconstruction of $F(T)$ and Invariant~Classification}

Using Morris--Thorne parametrization, the~effective shape function $b(r)$ can be reconstructed from the torsion-modified field equations, leading to
\begin{equation}
b_{\rm eff}(r) = r\left(1 - \frac{1}{A_2^2(r)}\right),
\end{equation}
which encodes the deviation from GR induced by nonlinear torsion. The~reconstruction requires the monotonicity condition
\begin{equation}
\frac{dT}{dr} \neq 0,
\end{equation}
which restricts admissible WH profiles. Physically viable models must satisfy
\begin{equation}
F_T > 0
\end{equation}
to ensure a positive effective gravitational coupling. This reconstruction generalizes the PL procedure developed in Sections~\ref{sect3} and \ref{sect4} to non-constant-radius~configurations.

{We adopt the invariant classification framework developed in Refs.~\cite{Landry2024_spherical,Landry2024_fluid,Landry2025_scalar,roberthudsonSSpaper}. The~classification of admissible $F(T)$ models follows the general structure introduced in \mbox{Sections~\ref{sect3} and \ref{sect4}}.} Here, we restrict attention to the subset of models 
compatible with WH-like geometries. The~existence of WH-like branches imposes additional constraints on the reconstruction procedure, namely the regularity of the torsion scalar at the throat and the absence of divergences in higher-order invariants~\cite{Ferraro2007,Bahamonde2023,Bahamonde2023b,Boehmer2011}. The~reconstruction requires \(T(r)\) to be locally invertible, at~least piecewise, in~the neighborhood of the throat. Equivalently,
\begin{equation}
\frac{dT}{dr}\neq0
\end{equation}
on the reconstruction branch under consideration, while critical points must be treated separately. This requirement ensures the invertibility of the mapping \(r\leftrightarrow T\), which is necessary for a consistent reconstruction of an \(F(T)\) solution.

Within the teleparallel invariant classification framework~\cite{Coley2020,Landry2024_spherical,Landry2024_fluid,Landry2025_scalar,roberthudsonSSpaper,Olver1995}, these models correspond to distinct invariant branches characterized by different scaling behaviors of the torsion invariants in Equation~\eqref{eqn46}. PL models may effectively mimic the stress-energy structure required at the throat, LOG corrections can soften the near-throat behavior, while EXP models may regularize torsion invariants but remain more sensitive to perturbations. Thus, Table~\ref{table3} should be interpreted as a local invariant classification of admissible WH-like branches, not as a proof that each branch gives a globally traversable WH. In~this case, WH-like configurations correspond to invariant subclasses where torsion scalars remain finite at the throat while supporting nontrivial geometric deformations.
	}

\begin{table}[ht]

\begin{tabular}{|c c c c c|}
	\toprule
	\textbf{Class} & \boldmath{\(F(T)\)} & \textbf{WH-like Branch} & \textbf{NEC Sector} & \textbf{Stability} \\
	\midrule
	TEGR & \(T\) & Exotic matter WH & Matter & GR-like \\
	PL & \(T+\alpha T^n\) & Possible & Effective torsion & Cond. stable \\
	Log & \(T+\alpha T\ln(T/T_0)\) & Possible smooth throat & Effective torsion & Near-throat stable \\
	Exp & \(T+\alpha e^{\lambda T}\) & Regular candidate & Model dependent & Sensitive \\
	Compo. & Mixed & Multi-scale candidate & Model dependent & Model dependent \\
	\bottomrule
\end{tabular}
	\caption{Local teleparallel invariant classification of teleparallel WH-like~branches.}
\label{table3}
\end{table}

{

	In contrast with GR, where traversable WHs require explicit exotic matter sources, nonlinear teleparallel $F(T)$ 	gravity provides an effective geometric mechanism through torsion corrections that can support such~configurations.

	The WH-like sector can be interpreted as a complementary 	branch of the $A_3 = r$ solutions discussed in Section~\ref{sect4}, which corresponds to configurations where horizon formation is avoided and the torsion sector regulates the core geometry. Although~WH-like geometries are often associated with regular cores, this is not automatic. The~behavior of torsion invariants must be examined explicitly to determine whether the geometry is truly regular or still contains hidden~singularities.

	{The distinction between BH-like and WH-like geometries can be summarized as~follows:}
\begin{equation}
	\mathrm{BH}: A_1^2(r_h)=0,\qquad
	\mathrm{WH}: A_1^2(r_0)\neq 0,\quad b(r_0)=r_0,\quad b'(r_0)<1.
	\end{equation}

\subsection{Singularity Structure and Energy~Conditions}

The singularity structure of WH-like geometries in \(F(T)\) gravity can be diagnosed using torsion invariants, in~addition to the usual metric and curvature regularity conditions. Regularity is assessed using the invariant set introduced in Section~\ref{sect25}. A~physically acceptable WH-like configuration requires not only finite torsion scalar \(T\), but~also finite higher-order invariants,
\begin{equation}
I_i < \infty,\quad i=1,\ldots,4,
\end{equation}
{which impose stronger constraints on admissible solutions. It is crucial to distinguish~between:}
\begin{enumerate}
	\item geometric regularity (finite torsion invariants),
	\item physical viability (ECs).
\end{enumerate}

In the covariant formulation, nonlinear torsion contributions can be recast as an 
effective energy-momentum tensor~\cite{Bahamonde2023,Bahamonde2023b,MorrisThorne1988}. The~null EC (NEC) can be decomposed into a matter contribution and 
an effective torsion contribution,
\begin{equation}
(\rho+P_r)_{\rm eff} = (\rho+P_r)_{\rm matter} + (\rho+P_r)_{\rm torsion}.
\end{equation}
WH-like configurations typically require an effective violation of the NEC, which may arise from the torsion sector rather than from the matter sector itself. This violation is branch-dependent and does not imply that standard matter necessarily violates classical~ECs.

This contrasts with the BH-like sector of Section~\ref{sect4}, where the central region is controlled primarily by the ultraviolet behavior of the exponent \(b\). The~torsion invariants determine whether the throat is genuinely regular or only formally horizonless. This distinction is important because the physical electromagnetic sector remains standard, while the effective torsion sector modifies the gravitational side of the field~equations.

\subsection{Stability Analysis and Physical~Interpretation}

The scalar-torsion stability conditions discussed in Sections~\ref{sect3} and \ref{sect4} remain necessary, but~are not sufficient for WH-like geometries. The~WH-like configurations introduce an additional constraint: the dynamical stability of the throat itself. Stability must therefore be analyzed at two distinct~levels:
\begin{enumerate}
\item Scalar-torsion~stability

\item Geometric stability of the throat.
\end{enumerate}

The stability of the throat can be analyzed by perturbing its radius $r_0$ and 
studying the resulting effective potential  \(V_{\rm eff}(r)\) as:
\begin{equation}
\left.\frac{d^2V_{\rm eff}}{dr^2}\right|_{r_0} > 0.
\end{equation}
Within the PL framework, a~representative, stable, and regularized branch is selected by
\begin{equation}
n=\frac{2a}{b+1}>1,\qquad b<-1,
\end{equation}
which combines scalar-torsion stability with softened throat~behavior.

WH--BH transition: in the charged case, the~RN limit is recovered when
\begin{equation}
e^{2\Phi(r)}\rightarrow 1-\frac{2M}{r}+\frac{Q^2+P^2}{r^2},
\end{equation}
and
\begin{equation}
b(r)\rightarrow 2M-\frac{Q^2+P^2}{r}.
\end{equation}

Viable WH-like solutions must satisfy a combination of regularity, stability, and~consistency conditions. These include finite torsion invariants, compatibility 
with antisymmetric field equations, positivity of the effective coupling (\(F_T>0\)), absence of tachyonic modes (\(F_{TT}>0\)), and~stability of the throat 
under radial perturbations~\cite{Boehmer2011,Bahamonde2023,Bahamonde2023b}.

From a physical perspective, these configurations may be interpreted as torsion-supported WH-like branches. Nonlinear \(F(T)\) corrections provide an effective geometric contribution to the throat structure and may mimic the stress-energy balance required to sustain WH-like configurations in restricted~branches.

Table~\ref{table4} summarizes possible WH-like branches but does not guarantee global traversability or full dynamical~stability.
\begin{table}[ht]
	\begin{tabular}{c c c c}
		\toprule
		\textbf{Class} & \textbf{Geometry} & \textbf{NEC Source} & \textbf{Stability} \\
		\midrule
		TEGR & Exotic matter WH & Matter sector & GR-like \\
		PL & Possible WH-like branch & Effective torsion sector & Cond. stable \\
		Log & Smooth near-throat branch & Effective torsion sector & Near-throat stable \\
		Exp & Regular WH candidate & Model dependent & Sensitive \\
		Compo & Multi-scale WH-like sector & Model dependent & Model dependent \\
		\bottomrule
	\end{tabular}
	\caption{{Classification} of wormhole-like branches in nonlinear \(F(T)\) gravity.}
	\label{table4}
\end{table}

Overall, WH-like solutions require a delicate balance between torsion dynamics, electromagnetic contributions, and~geometric constraints. Only restricted nonlinear \(F(T)\) branches can simultaneously satisfy throat regularity, finite torsion invariants, compatibility with the antisymmetric field equations, positive effective coupling, and~dynamical throat~stability.

\section{Discussion and~Conclusions}\label{sect6}

In this work, we have presented a covariant analysis of static SS configurations in \(F(T)\) gravity coupled to Maxwell fields, covering both the constant-radius sector \(A_3=c_0\) and the areal-radius sector \(A_3=r\). Starting from the CSC formalism and the associated symmetric and antisymmetric field equations, we showed that the electromagnetic sector remains strongly constrained by the covariant teleparallel structure. In~particular, the~antisymmetric field equations restrict admissible CSC pairs and favor radial electric and magnetic configurations compatible with the standard Maxwell scaling~\cite{Krssak2015,Golovnev2020}. In~the constant-radius regime, the~electromagnetic sector behaves effectively as a cosmological source, leading to Nariai-type and vacuum-dominated branches~\cite{Cai2016,Bahamonde2023,Bahamonde2023b}. By~contrast, the~\(A_3=r\) sector contains the physically relevant charged compact-object configurations, including RN-like geometries and nonlinear \(F(T)\) deformations, where torsion corrections modify the horizon structure, near-core behavior, and~asymptotic properties~\cite{Awad2017,Nashed2019}.

A central result of the present analysis is that nonlinear teleparallel gravity enlarges the space of admissible charged solutions while preserving the standard gauge structure of the Maxwell sector. The~reconstruction procedure developed for PL coframes provides explicit \(F(T)\) branches, including PL, LOG, EXP, and~COMP models, which can be organized within the teleparallel invariant classification framework~\cite{Landry2024_spherical,Landry2024_fluid,Landry2025_scalar,roberthudsonSSpaper}. The~lapse-function profiles and torsion-scalar analysis show that the parameters controlling the reconstructed models determine horizon formation, singularity behavior, and~possible regularization. In~particular, the~critical regimes \(b>-1\), \(b=-1\), and~\(b<-1\) respectively distinguish singular, critical, and~regularized torsion-core behavior. The~AdS-like charged branches further demonstrate how an effective cosmological term can be incorporated into the teleparallel compact-object~sector.

The WH-like sector complements the BH-like solutions by replacing horizon formation with throat formation. Nonlinear torsion contributions may effectively support the flaring-out condition and shift the NEC balance into the geometric sector, while the physical electromagnetic stress-energy tensor remains standard~\cite{Cai2016,MorrisThorne1988}. However, this mechanism is branch-dependent and does not imply that all nonlinear \(F(T)\) models generate physically viable traversable WHs. Full viability requires throat regularity, finite torsion invariants, compatibility with the antisymmetric field equations, positive effective coupling, absence of tachyonic modes, and~dynamical stability of the throat under radial perturbations. Thus, the~WH-like branches should be interpreted as admissible local geometric sectors rather than automatically globally traversable solutions. This perspective is further supported by recent work in covariant teleparallel gravity, where explicit solutions of the field equations have been used to reconstruct viable \(F(T)\) models describing weak-massive WH configurations. This further illustrates how torsion contributions can act as effective geometric sources in compact-object and WH-like sectors~\cite{LandryWHmass2026}.

Finally, the~stability analysis indicates that physically viable reconstructed models must satisfy \(F_T>0\) and \(F_{TT}>0\), ensuring the absence of ghost-like and tachyonic scalar-torsion modes. LOG and selected PL branches appear to be the most robust, while EXP and COMP models remain more sensitive to perturbations. From~an observational perspective, nonlinear torsion corrections may lead to deviations in BH shadows, gravitational lensing, photon-sphere structure, and~quasi-normal mode spectra~\cite{Bahamonde2023,Bahamonde2023b,Awad2017,Nashed2019}. Future work should address coupled perturbations, geodesic completeness, global WH traversability, and~strong-field observational constraints in order to constrain possible torsion signatures in the gravitational~sector.

}

\vspace{6pt} 

\section*{Acknowledgements}

Thanks to A. A. Coley for his constructive comments.



%


\end{document}